\documentclass{aa}
\usepackage{times,float}
\usepackage{epsfig}
\usepackage{graphics}
\usepackage{amssymb}
\usepackage{curves}
\usepackage{eepic}
\usepackage{epic}

\begin{document}

\title{The GRB~030329 host: a  blue low metallicity subluminous galaxy with
intense star  formation\thanks{Based on  observations made with  the Nordic
Optical Telescope, operated  on the island of La  Palma jointly by Denmark,
Finland, Iceland, Norway, and Sweden, in the Spanish Observatorio del Roque
de los Muchachos of the Instituto de Astrof\'{\i}sica de Canarias. Based on
data taken  at the 2.2-m and  3.5-m telescopes of  the Centro Astron\'omico
Hispano Alem\'an  de Calar  Alto, operated by  the Max Planck  institute of
Heidelberg  and Centro Superior  de Investigaciones  Cient\'{\i}ficas.  The
spectral observations  were obtained at the  European Southern Observatory,
Cerro Paranal  (Chile), under  the Director's Discretionary  Time programme
271.D-5006(A).}}

\titlerunning{The host galaxy of GRB~030329}

\author{
        J.~Gorosabel\inst{1}
   \and D.~P\'erez-Ram\'\i rez\inst{2}
   \and J.~Sollerman\inst{3}
   \and A.~de Ugarte Postigo\inst{1}
   \and J.~P.~U.~Fynbo\inst{4}
   \and A.~J.~Castro-Tirado\inst{1}
   \and P.~Jakobsson\inst{4,5}
   \and L.~Christensen\inst{6}    
   \and J.~Hjorth\inst{4}
   \and G.~J\'ohannesson\inst{5}
   \and S.~Guziy\inst{1}
   \and J.~M.~Castro Cer\'on\inst{4}
   \and G.~Bj\"ornsson\inst{5}
   \and V.~V.~Sokolov\inst{7}
   \and T.~A.~Fatkhullin\inst{7}
   \and K. Nilsson\inst{4,8}
}

\institute{ 
           Instituto de Astrof\'{\i}sica de Andaluc\'{\i}a (IAA-CSIC),
           P.O. Box 03004, E-18080 Granada, Spain; {\tt jgu@iaa.es,
            ajct@iaa.es, deugarte@iaa.es, gss@iaa.es}
           \and
           Universidad de Ja\'en, Departamento de F\'\i sica (EPS),
           Virgen de la Cabeza, 2, Ja\'en, E-23071, Spain;
           {\tt dperez@ujaen.es}
           \and
           Stockholm Observatory, Department of Astronomy, AlbaNova, S-106
           91 Stockholm, Sweden; {\tt jesper@astro.su.se}
           \and
           Niels Bohr  Institute, University of  Copenhagen, Juliane Maries
           Vej  30, 2100  Copenhagen \O,  Denmark;  {\tt 
           jfynbo@astro.ku.dk, pallja@astro.ku.dk, jens@astro.ku.dk,
           josemari@alumni.nd.edu, kim@astro.ku.dk}
           \and
            Science Institute, University of Iceland, Dunhaga 3, 107
            Reykjav\'{\i}k, Iceland; {\tt gudlaugu@raunvis.hi.is, 
            gulli@raunvis.hi.is}
           \and
            Astrophysikalisches Institut, 14482 Potsdam, Germany;
            {\tt lchristensen@aip.de}
            \and
            Special Astrophysical Observatory of the Russian Academy of
            Sciences, Nizhnij Arkhyz, 357 147 Karachai-Cherkessia, Rusia;
            {\tt sokolov@sao.ru, timur@sao.ru}
            \and
            European  Southern  Observatory, Karl-Schwarzschild-Strasse  2,
            D-85748 Garching bei Munchen, Germany.
           }

\offprints{ \hbox{J. Gorosabel, e-mail:{\tt jgu@iaa.es}}}

\date{Received  / Accepted }


\abstract{We   present  broad   band   photometry  and   spectroscopic
  observations  of the  host galaxy  of GRB~030329.   Analysis  of the
  spectral  emission  lines  shows  that  the host  is  likely  a  low
  metallicity galaxy ($Z\sim0.004$).  The spectral energy distribution
  (SED) constructed with the  photometric points has been fitted using
  synthetic and observational templates.  The best SED fit is obtained
  with  a starburst  template  with an  age  of $\sim150$  Myr and  an
  extinction $A_{\rm v}  \sim 0.6$.  We find that  the GRB~030329 host
  galaxy is  a subluminous  galaxy ($L \sim  0.016 L^{\star}$)  with a
  stellar  mass  of  $\gtrsim  10^{8} M_{\odot}$.   Three  independent
  diagnostics, based on the restframe UV continuum, the [\ion{O}{ii}],
  and the Balmer emission lines, provide a consistent unextincted star
  formation rate of $ \sim  0.6 M_{\odot} $ yr$^{-1}$, implying a high
  unextincted  specific  star  formation  rate  ($\sim  34  M_{\odot}$
  yr$^{-1}  (L/L^{\star})^{-1}$).  We  estimate  that the  unextincted
  specific star formation  rate of the GRB~030329 host  is higher than
  $\sim93.5\%$ of the galaxies at a similar redshift.   }

\maketitle

\section{Introduction}

An extremely bright gamma--ray burst  (GRB) was detected on 2003 March
29 at 11:37:14.67 UT (GRB~030329; Vanderspek et al.  \cite{Vand03}) by
the French Gamma Ray Telescope (FREGATE), the Wide Field X-Ray Monitor
(WXM), and  the Soft X-Ray  Camera (SXC) instruments aboard  the $High
~Energy ~Transient  ~Explorer ~II$ (HETE-2) spacecraft  (Ricker et al.
\cite{Rick03}).  With a duration  of approximately 30~s in the 30--400
keV energy range, GRB~030329 falls into the ``long-duration'' category
of GRBs (Kouveliotou et al. \cite{Kouv93}).

Before  the HETE-2 era,  two main  scenarios were  favoured to  explain the
nature  of long  GRBs.  One  of  them suggested  that GRBs  occur when  two
compact objects  (such as black holes  or neutron stars)  merge (Eichler et
al.  \cite{Eich89};  Narayan et al.  \cite{Nara92}). This  would imply that
GRBs occur  after massive star formation  has ended in a  galaxy, since the
time-scale for binary objects to  merge is long ($\gtrsim1$~Gyr).  A second
scenario,  the  so-called  collapsar  scenario,  suggested  that  GRBs  are
associated  with   the  core-collapse  of  massive   stars  (supernovae  or
hypernovae: Woosley \cite{Woos93};  Paczy\'nski \cite{Pacz98}; MacFadyen \&
Woosley   \cite{MacF99}).   This   scenario  predicts   a   supernova  (SN)
accompanying the GRB,  which thus must be coincident with the epoch of
star formation in the host.

The  studies   carried  out   in  2003  for   GRB~030329  (Hjorth   et  al.
\cite{Hjor03}; Stanek et al.   \cite{Stan03}; Sokolov et al. \cite{Soko03};
Kawabata  et  al.   \cite{Kawa03})  strongly confirmed  previous  evidences
(Bloom  et al.   \cite{Bloo99}; Castro-Tirado  \&  Gorosabel \cite{Cast99})
that long GRBs are physically linked to supernova explosions. The origin of
the short GRBs is still unknown, so  here we will discuss only the hosts of
long GRBs. Radio, optical and/or infrared afterglows have been observed for
$\sim$70 long GRBs,  and the  majority of  these coincide  with star
forming galaxies.

  The  pioneering  statistical   work  applying  the  spectral  energy
distribution (SED) fitting technique to  GRB host galaxies was published by
Sokolov et  al.  (\cite{Soko01}), who  argued that GRB hosts  correspond to
starburst  galaxies.  A  number of  other complementary  studies  have been
carried  out in the  past, studying  the properties  of different  GRB host
samples  (Chary et  al.  \cite{Char02};  Berger et  al.   \cite{Berg03}; Le
Floc'h et al.  \cite{LeFl03}; Tanvir et al.  \cite{Tanv04}) or centred on a
particular host  galaxy (Djorgovski  et al.  \cite{Djor01a};  Djorgovski et
al.    \cite{Djor03};  Hjorth   et  al.    \cite{Hjor02}).    Recently,  a
statistical  analysis  by Christensen  et  al.  (\cite{Chri04a};  hereafter
C04A) based on  the SED of a  sample of 10 GRB host  galaxies indicated the
similarity of  GRB hosts  and the young  starburst galaxies present  in the
Hubble Deep Field (HDF; Williams et al. \cite{Will00}).

A good knowledge  of the host SED makes it possible  to derive the key
parameters  for determining the  conditions that  favour a  GRB event,
e.g., the  star formation rate (SFR), the  dominant stellar population
age,   the stellar  mass and the  extinction.  GRB  host galaxies
tend to be  faint ($R > 23$;  Djorgovski et al. \cite{Djor01b}),
and thus,  spectroscopic studies of  the SED require extensive  use of
large telescopes.   An economic alternative that allows  us to extract
information  on  the properties  of  the  host  galaxies is  based  on
multicolour  broad band  imaging.  The main  inconvenience of  fitting
templates to  photometric points comes  from the insensitivity  of the
SED  solution   to  the   assumed  metallicity  (Christensen   et  al.
\cite{Chri04b}),  which   is  considered  as   a  secondary  parameter
(Bolzonella  et  al.   \cite{Bolz00}).    In  practice,  in  order  to
determine  reliable  metallicities for  the  GRB hosts,  spectroscopic
observations are therefore required.

The         optical         afterglow         (OA)         coordinates
($\alpha_{J2000}=10^{h}44^{m}59\fs95,
\delta_{J2000}=+21^{\circ}31^{\prime}17\farcs8$ with an uncertainty of
0.5  arcsec in  each axis;  Price  et al.   \cite{Pric03}) allowed  an
extensive follow up from the Northern hemisphere, although it was also
visible from the main  observatories of the Southern hemisphere.  The
redshift of GRB~030329 was determined  to be $z=0.168$ (Greiner et al.
\cite{Grei03}) from  high-resolution spectroscopy with  the Very Large
Telescope  (VLT).   This  makes  GRB~030329 the  third  nearest  burst
overall  (GRB~980425  is the  nearest  at  $z=0.0085$,  Galama et  al.
\cite{Gala98};  and  GRB~031203   had  $z=0.1055$,  Prochaska  et  al.
\cite{Proc04}).  As the OA faded, prominent spectral features revealed
a Type  Ic supernova (named SN~2003dh) peaking 10--15  days after the
GRB event (Hjorth et al.  \cite{Hjor03}; Stanek et al. \cite{Stan03}).

\begin{table*}
\begin{center}
\caption{Optical and NIR photometric observations carried out for the GRB~030329 host galaxy.}
\begin{tabular}{lccccc}
\hline
\hline
Telescope     &  Filter & Date UT &$t_{\rm exp}$&  Seeing  & Limiting magnitude \\
(+Instrument) &         &         &   (s)       & (arcsec) &    ($3\sigma$)     \\
\hline
NOT     (+MOSCA) &  $U$&  24.03--24.08/03/04&   5$\times$900& 1.0 & 25.4\\
NOT     (+MOSCA) &  $B$&  24.85--24.89/03/04&   3$\times$600& 1.1 & 25.8\\
NOT     (+MOSCA) &  $V$&  24.86--24.87/03/04&   1$\times$600& 1.0 & 24.4\\
NOT     (+MOSCA) &  $R$&  24.87--24.88/03/04&   1$\times$600& 0.9 & 24.2\\
2.2-m    (+BUSCA)& $C1$&  20.94--21.01/03/04&  19$\times$600& 1.8$^{\dagger}$& 25.6$^{\star}$\\
                 &     &  21.97--22.12/03/04&  14$\times$600&     &     \\
                 &     &  22.95--23.06/03/04&  12$\times$600&     &     \\
2.2-m    (+BUSCA)& $C2$&  20.94--21.01/03/04&  19$\times$600& 1.7$^{\dagger}$& 25.1$^{\star}$\\
                 &     &  21.97--22.12/03/04&  14$\times$600&     &     \\
                 &     &  22.95--23.06/03/04&  12$\times$600&     &     \\
2.2-m    (+BUSCA)& $C3$&  20.94--21.01/03/04&  19$\times$600& 1.7$^{\dagger}$& 25.2$^{\star}$\\
                 &     &  21.97--22.12/03/04&  14$\times$600&     &     \\
                 &     &  22.95--23.06/03/04&  12$\times$600&     &     \\
2.2-m    (+BUSCA)& $C4$&  20.94--21.01/03/04&  19$\times$600& 1.6$^{\dagger}$& 24.1$^{\star}$\\
                 &     &  21.97--22.12/03/04&  14$\times$600&     &     \\
                 &     &  22.95--23.06/03/04&  12$\times$600&     &     \\
3.5-m    (+Omega-Prime) &  $J$&  05.07--05.19/01/04& 113$\times$60 & 0.7 & 24.1\\
3.5-m    (+Omega-Prime) &  $H$&  06.07--06.18/01/04& 109$\times$60 & 0.8 & 22.9\\
3.5-m    (+Omega-Prime) &  $K^{\prime}$&  07.06--07.20/01/04&  99$\times$60 & 0.8 & 19.7\\
\hline
\multicolumn{6}{l}{$\dagger$ Mean seeing of the co-added image.}\\
\multicolumn{6}{l}{$\star$   Limiting magnitude obtained co-adding the three epochs.}\\
\hline             
\label{table1}
\end{tabular}
\end{center}
\end{table*}

\begin{table*}
\begin{center}
\caption{Optical/NIR  magnitudes of  the  host.  The  fourth  and
fifth columns  show the  measured magnitudes in  the Vega  and AB
systems (not corrected for Galactic reddening).}
\begin{tabular}{lcccc}
\hline
\hline
Filter name          & Effective       & Bandpass    &Magnitude&  Magnitude\\
(Instrument)         & wavelength (\AA)& width (\AA) &  (Vega) &   (AB)   \\
\hline                                                                      
 $U$ (MOSCA)         &   3638.3 &  396.3      & 22.72$\pm$0.10 & 23.45$\pm$0.10\\
 $C1$ (BUSCA)        &   3654.5 &  1081.6     & 23.03$\pm$0.03 & 23.51$\pm$0.03\\ 
 $B$ (MOSCA)         &   4358.7 &  680.5      & 23.33$\pm$0.07 & 23.26$\pm$0.07\\
 $C2$ (BUSCA)        &   5007.4 &  706.2      & 22.92$\pm$0.05 & 22.87$\pm$0.05\\ 
 $V$ (MOSCA)         &   5413.4 &  636.4      & 22.83$\pm$0.10 & 22.84$\pm$0.10\\
 $R$ (MOSCA)         &   6422.1&  1072.5      & 22.66$\pm$0.04 & 22.86$\pm$0.04\\
 $C3$ (BUSCA)        &   6455.8 & 1147.0      & 22.61$\pm$0.04 & 22.81$\pm$0.04\\
 $C4$ (BUSCA)        &   7999.7 &  780.7      & 22.10$\pm$0.04 & 22.55$\pm$0.04\\
 $J$ (Omega-Prime)        &  12861.5 & 1713.9 & 21.45$\pm$0.16 & 22.42$\pm$0.16\\
 $H$ (Omega-Prime)        &  16491.6 & 1814.2 & 21.15$\pm$0.24 & 22.55$\pm$0.24\\
 $K^{\prime}$(Omega-Prime)&  21634.9 & 1636.5 & $>$19.70$\dag$ &$>$ 21.57$\dag$\\
\hline
\multicolumn{5}{l}{$\dag$ 3 $\sigma$ upper limit}\\
\hline
\label{table2}
\end{tabular}
\end{center}
\end{table*}

In  this  paper  we  present  a  comprehensive  multicolour  study  of  the
GRB~030329  host galaxy,  similar  to those  performed  for the  GRB~000210
(Gorosabel   et  al.    \cite{Goro03a}),  GRB~000418   (Gorosabel   et  al.
\cite{Goro03b})  and GRB~990712 (Christensen  et al.   \cite{Chri04b}) host
galaxies. The  aim of the  analysis is to  determine the properties  of the
stellar populations  dominating the optical/near-infrared  (NIR) light from
the  host  galaxy  and  the  amount  of  extinction  due  to  dust  in  the
interstellar medium (ISM) of the host.

Given  the faintness  of GRB host  galaxies in  general, detailed
observations of the low redshift hosts, as in the present study, might
be  used to  construct  a library  of  spectral templates,  applicable
(under several  assumptions) to the hosts placed  at higher redshifts.
Furthermore, the case of  GRB~030329 is particularly remarkable thanks
to its association with  SN~2003dh.  Both arguments justify a detailed
analysis of this host galaxy.

Throughout    this    paper,    we    assume   a    cosmology    where
$\Omega_{\Lambda}=0.7$,  $\Omega_{M}=0.3$  and  $H_0=65$  km  s$^{-1}$
Mpc$^{-1}$.   Under  these  assumptions,  the luminosity  distance  of
GRB~030329 is $d_l=872.5$ Mpc and the look-back time is 2.25~Gyr (15.5
\% of the present Universe age).

Sect.~\ref{Observations}   describes  the  photometric   and  spectroscopic
observations carried out  for the GRB~030329 host galaxy.  Our main results
are reported in  Sect.~\ref{results}.  In Sect.~\ref{Discussion} we discuss
the implications  of   our work and  Sect.~\ref{Conclusions} summarises
the final conclusions.

\section{Observations and Analysis}
\label{Observations}
\subsection{Imaging}

We have used a number of  optical/NIR facilities in order to compile a well
sampled SED  (see Table~\ref{table1}). The $UBVR$-bands  were observed with
the 2.56-m Nordic Optical Telescope (NOT) equipped with MOSCA, a 2$\times$2
mosaic  2048$\times$2048 pixel  CCD,  covering  a field  of  view (FOV)  of
$7\farcm7  \times 7\farcm7$.   These  observations were  carried  out in  a
2$\times$2 binning mode, yielding a pixel scale of $0\farcs22$/pix.

\begin{figure}[t]
\begin{center}
  {\includegraphics[width=\hsize]{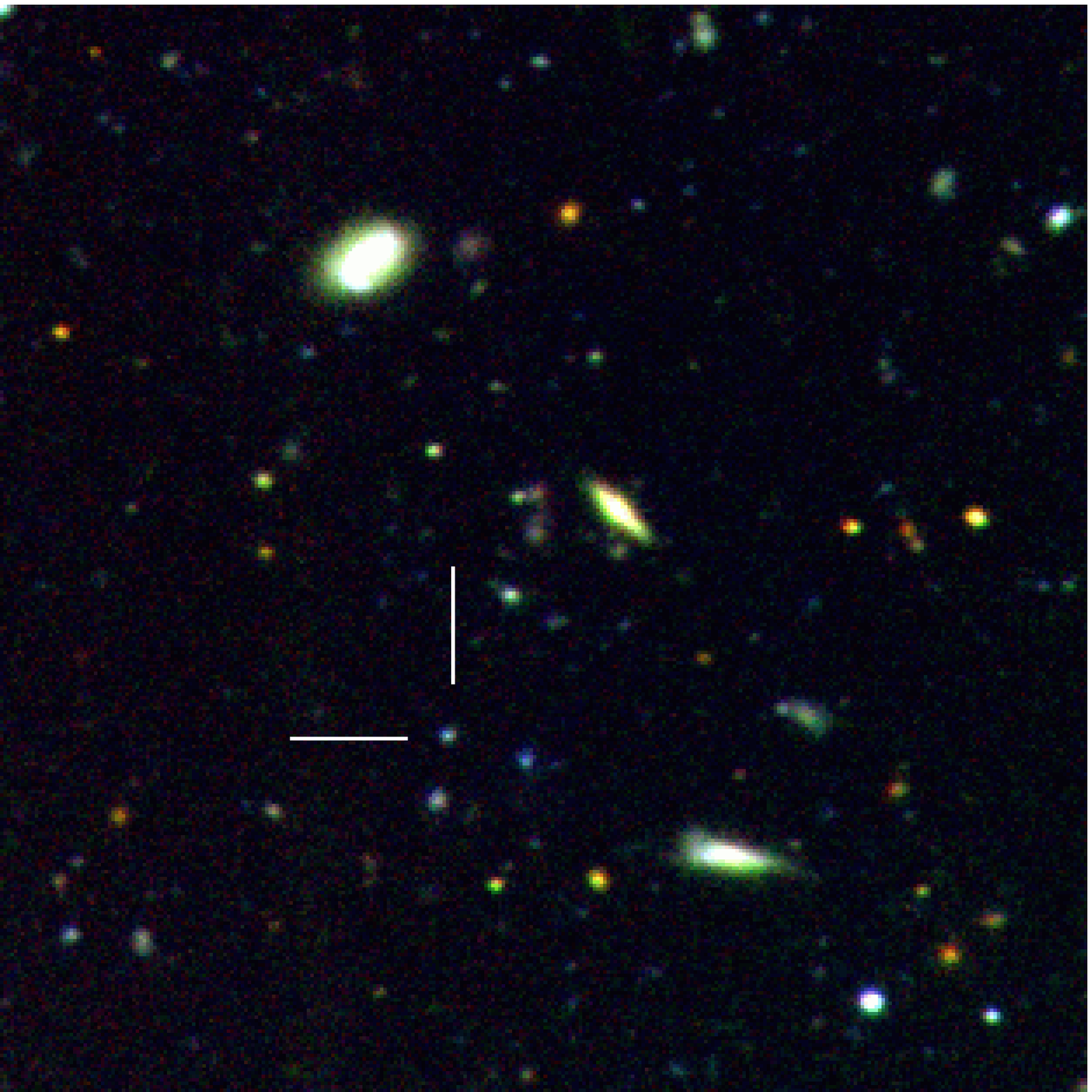}}
  \caption{\label{fig1}  The  field of  the  GRB~030329  host galaxy.   The
  picture is  a coloured composite  image created by combining  three BUSCA
  images taken  in the $C1$,  $C3$, and $C4$  bands.  The FOV  is $2\farcm0
  \times 2\farcm0$.  The location of  the host galaxy is indicated.  As can
  be  seen from  the picture,  the  host shows  a bluish  colour. North  is
  upwards and East is leftwards.}
\end{center}
\end{figure}

Additional optical  observations were obtained with the  BUSCA camera (Reif
et   al.     \cite{Reif00})   at    the   2.2-m   telescope    located   at
CAHA\footnote{Centro   Astron\'omico  Hispano   Alem\'an}.    BUSCA  allows
simultaneous  imaging  in four  broad  optical  bands.   The four  channels
(hereafter named  $C1$, $C2$, $C3$ and  $C4$) are not  standard filters and
have  to  be calibrated  by  observing  spectro-photometric standard  stars
(characteristics  of  the  four  BUSCA  channels  are  described  in  Table
\ref{table2}).  $C1$  resembles the Johnson $U$-band, $C2$  is a transition
between  Johnson's  $B$  and  $V$  filters.   $C3$ is  very  close  to  the
$R_c$-band  and   $C4$  is  basically   identical  to  the   $I_c$  filter.
Figure~\ref{fig1} shows  a coloured composite  image based on three  out of
four BUSCA channels,  namely $C1$, $C2$, and $C4$.   BUSCA consists of four
4K$\times$4K pixel CCDs plus  3 dichroics.  The observations were performed
in  the 2$\times$2  binning mode.   The  FOV covered  is $12\farcm0  \times
12\farcm0$, and the resulting pixel scale was $0\farcs35$/pix.

Finally, the  $JHK^{\prime}$ observations were  carried out with  the 3.5-m
CAHA telescope  equipped with Omega-Prime, a  1024$\times$1024 pixel HgCdTe
CCD, which provides  a FOV of $6\farcm8 \times 6\farcm8$  and a pixel scale
of $0\farcs39$/pix.

Reduction  was  performed   following  standard  procedures  running  under
IRAF\footnote{IRAF  is  distributed   by  the  National  Optical  Astronomy
Observatories, which  are operated by  the Association of  Universities for
Research in Astronomy, Inc.,  under cooperative agreement with the National
Science Foundation.}.   In Table \ref{table1} we provide  the observing log
for our optical and NIR imaging observations.

The optical/NIR  imaging was carried out  at least 282  days after the
GRB,  when the OA+SN  contamination was  negligible.  The  host galaxy
appears  as a  star-like object  in all  our images,  showing  NIR and
optical  profiles  consistent with  those  of  field  stars.  This  is
consistent   with   the   HST   observations  by   Fruchter   et   al.
(\cite{Fruc03}), who reported  that the host is a  dwarf galaxy with a
brightness  of   $V=22.7  \pm  0.3$  (consistent   with  our  $V$-band
magnitude, see Table ~\ref{table2}).   According to the angular scales
of the OA-host system discussed in Fruchter et al.  (\cite{Fruc03}), a
host  compact  appearance  is  expected for  seeing  conditions  above
$\sim0\farcs5$,  in  agreement  with  our observations  (seeing  $\geq
0\farcs7$; see  Table~\ref{table1}).  Therefore, considering  that the
relative photometry  is independent on  the aperture radius,  the host
magnitudes given in Table \ref{table2}, are based on circular aperture
measurements (PHOT  running under  IRAF) with no  aperture corrections
and an  aperture radius of  1.5 times the  full width at  half maximum
(FWHM).

The  $UBVR$-band calibration is  based on  Henden (\cite{Hend03}),  and the
$JHK^{\prime}$ calibration on Guziy et al.  (\cite{Guzi05}).  For the BUSCA
data,   the   flux  calibration   was   carried   out   by  observing   the
spectro-photometric standard  star GD153 (Bohlin et  al.  \cite{Bohl95}) at
the same  airmass to  that of the  GRB host  on a photometric  night.  This
allowed us  to establish secondary field  standard stars in  the AB system.
As can be  seen in Fig.~\ref{fig2}, the agreement  of the BUSCA calibration
with the NOT optical magnitudes is satisfactory.

\subsection{Spectroscopy}

Spectroscopy of  the host of GRB 030329  was obtained on 2003  June 19 with
 the    Focal   Reducer   and    Low   Dispersion    Spectrograph   (FORS2)
 instrument\footnote{http://www.eso.org/instruments/fors/}  attached to the
 8.2-m UT4 of  the European Southern Observatory Very  Large Telescope. The
 FORS2 instrument  is equipped with two  $2048 \times 4096$  pixel MIT CCDs
 having $15~\mu$m pixels.  We used  the standard resolution mode that gives
 a FOV of $6\farcm8 \times 6\farcm8$,  and with $2 \times 2$ binning on the
 chip we obtained a pixel size on the sky of $0\farcs25$.

\begin{figure*}[t]
\begin{center}
\resizebox{\hsize}{!}{\includegraphics{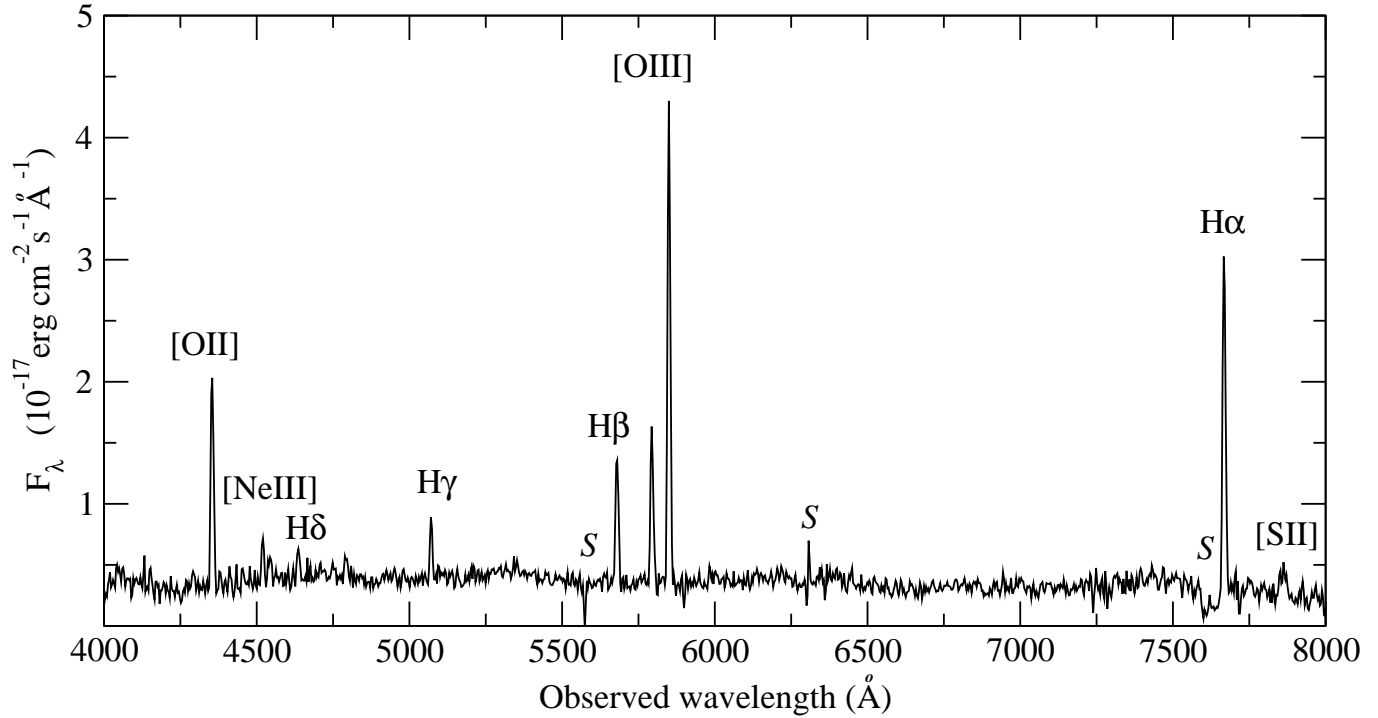}}
\caption{\label{spec} The  GRB\,030329 host galaxy  VLT spectrum.  The
  flux   is   corrected   for  Galactic   reddening   ($E(B-V)=0.025$)
  considering a  MW extinction  law (Cardelli et  al.  \cite{Card89}).
  The  residuals resulting  from  the telluric  line subtractions  are
  indicated by an $S$.  Note the telluric line affecting the left wing
  of  H$\alpha$.  The spectrum  was acquired  $\sim82$ days  after the
  GRB, when the  contribution to the continuum by  the OA+SN was still
  significant.}
\end{center}
\end{figure*}

 The  observations were  conducted with  the 300V  grism and  an order
 sorting filter  GG375, which efficiently covers  the wavelength range
 from $\sim 3800$--$8800$~\AA.  A 1\farcs3  wide slit was used and the
 seeing during the observations was typically below 0\farcs6, yielding
 a spectral  resolution of $\sim  10.5$~\AA.  Given the small  size of
 the  object,  the entire  flux  of the  host  galaxy  should thus  be
 included in the  slit. A position angle of 123.6  degrees was used to
 align  also   a  nearby  star  on   the  slit  (see   Hjorth  et  al.
 \cite{Hjor03}).   Three spectra  of  900~s each  were obtained,  with
 small  offsets on  the slit  between the  individual  exposures.  The
 airmass was  always quite  substantial since these  observations were
 obtained from  the Southern hemisphere,  and varied from 1.8  to 2.3.
 The final combined spectrum is shown in Fig.~\ref{spec}.

The spectra  were reduced  in a standard  way, including  bias subtraction,
flat fielding,  and wavelength calibration using spectra  of a Helium-Argon
lamp.   Flux  calibration  was  done  relative  to  the  spectrophotometric
standard star Feige67 (Oke \cite{Oke90}).  The absolute flux calibration of
the combined spectrum was obtained relative to our broad band photometry.

We note  that our host  galaxy spectroscopic observations were  carried out
$\sim82$ days  after the GRB,  when the contribution  of the OA+SN  and the
host were comparable  (Guziy et al. \cite{Guzi05}).  In  contrast, the host
properties derived by Matheson et al.  (\cite{Math03}) are based on spectra
acquired only $6-10$  days after the GRB, when the  OA+SN was still $50-25$
times  brighter than the  host.  This  intense overimposed  radiation field
might affect the measurement of  the host emission lines fluxes (especially
the faint  ones), so we consider our  spectrum a cleaner probe  of the host
galaxy physical properties.

Before  using the  emission  lines  as indicators  of  several host  galaxy
properties, they  were corrected for  Galactic foreground reddening  in the
direction  of  the GRB  ($E(B-V)=0.025$;  Schlegel  et al.   \cite{Schl98})
considering a MW extinction law (Cardelli et al.  \cite{Card89}).

\subsection{Photometry: Constructing the optical/NIR SED}
\label{method}

In  order to  derive the  corresponding effective  wavelengths  and AB
magnitudes  we  convolved  each  filter transmission  curve  plus  the
corresponding CCD  efficiency curve (see Table  \ref{table2}) with the
Vega  spectrum ($\alpha$  Lyrae,  $m=0$ in  all  bands by  definition;
Bruzual \& Charlot \cite{Bruz93}). The  fit of the SED has been mainly
obtained  using  synthetic  templates,   but  checks  have  also  been
performed using observational  templates (Kinney et al. \cite{Kinn96},
hereafter K96).

The     synthetic    SED     analysis    is     based    on     the    code
HyperZ\footnote{http://webast.ast.obs-mip.fr/hyperz/}  (Bolzonella  et  al.
\cite{Bolz00}).   Eight  synthetic spectral  types  were used  representing
Starburst galaxies  (Stb), Ellipticals (E), Lenticulars  (S0), Spirals (Sa,
Sb,  Sc  and Sd)  and  Irregular galaxies  (Im).   Each  spectral type  has
associated  a SFR($t$) temporal  history, and  hence a  characteristic time
scale (SFR  $\propto \exp{(-t/\tau)}$;  where $\tau$ can  range from  0 for
Stb, to  $\infty$ for  Im galaxies).  For  the generation of  the synthetic
templates, two initial mass functions (IMFs) have been considered; Salpeter
(\cite{Salp55}; S55), and  Chabrier (\cite{Chab03}; CH03).  Four extinction
laws  have  been  taken  into  account: Calzetti  et  al.   (\cite{Calz00};
suitable  for  Stbs),  Seaton  (\cite{Seat79};  for  the  Milky  Way,  MW),
Fitzpatrick  (\cite{Fitz86};  for the  Large  Magellanic  Cloud, LMC),  and
Pr\'evot et al.  (\cite{Prev84}; for  the Small Magellanic Cloud, SMC). The
construction of the  HyperZ templates has been performed  using the GALAXEV
public  code  (Bruzual  \&  Charlot  \cite{Bruz03}),  using  as  input  the
metallicity  ($Z$,  see   Sect.~\ref{metallicity})  derived  from  our  VLT
spectroscopic observations.

In addition to the aforementioned evolutionary templates, the observational
templates from  K96 were considered.   These extra spectral  templates give
complementary information on the  galaxy type and extinction.  The observed
K96 templates can be grouped into seven  sets: Bulge (B), E, S0, Sa, Sb, Sc
and Stb galaxies.  Additionally, the  Stb K96 templates are subdivided into
six classes depending  on their extinction by increasing  $E(B-V)$ in steps
of 0.1 (from Stb1 to Stb6; see more details in K96). The extinction for the
B, E, S0, Sa, Sb, and Sc templates is a free parameter.

For  the  construction of  the  photometric  SED  all the  magnitudes  were
corrected for  Galactic foreground  reddening along the  GRB line  of sight
(Schlegel  et al.   \cite{Schl98})  assuming a  typical  MW extinction  law
(Cardelli et al. \cite{Card89}).

\section{Results}
\label{results}

The VLT spectrum allows us  to estimate the host galaxy extinction ($A_{\rm
v}$), the  metallicity ($Z$) and  the SFR, based  on the oxygen  and Balmer
emission lines.   Table~\ref{table1c} displays  the list of  emission lines
detected in our VLT spectrum. The  flux of the lines, with the exception of
H${\delta}$, are  consistent within 1.5$\sigma$ with  the fluxes previously
reported for the GRB 030329 host galaxy (Hjorth et al. \cite{Hjor03}).

Alternatively,  the SED constructed  from the  optical/NIR photometric
points gives  an independent estimate  of the extinction and  the SFR.
The SED  also provides us  with information on the  stellar population
age, the  stellar mass,  and on the  host galaxy  absolute luminosity.
For self-consistency,  we based our SED fitting  analysis on templates
constructed  adopting the metallicity  derived from  our spectroscopic
study ($Z=0.004$). Table~\ref{table2} shows the optical/NIR magnitudes
given in the  Vega and AB systems, not corrected  for Galactic or host
reddening.   The listed  magnitude errors  account for  both  the zero
point  and  the   statistical  errors.   Figure~\ref{fig2}  shows  our
photometric points and the SED  solution obtained for a SMC extinction
law, a  S55 IMF  and a metallicity  of $Z=0.004$. Considering  all the
possible extinction laws  (4) and IMFs (2), we end  up with 8 possible
templates for each galaxy type.

\subsection{Photometric redshift}
\label{photoz}

Given that the spectroscopic redshift is known, we only use the photometric
redshift  to   check  our  fitted  SEDs  for   internal  consistency.   The
spectroscopic  redshift falls  within the  68\% uncertainty  region  of the
photometric redshift in 6 out of the 8 IMF/extinction law combinations.

A  rough estimate  based on  the  K96 templates  yields a  photometric
redshift around  $z\sim 0.11$,  reasonably close to  the spectroscopic
redshift.   We  consider  the  inferred photometric  redshifts  to  be
basically consistent with the spectroscopic redshift, and hereafter we
fix the redshift of the SEDs to $z=0.168$.

\subsection{Absolute magnitude}
\label{subluminous}  

Subluminous galaxies are defined as having luminosity values below the knee
of  the  luminosity  function  given  by  $L^{\star}$  (or  the  equivalent
$M^{\star}_B$).   Slightly  different values  for  $M^{\star}_B$ have  been
reported  depending  on  the   galaxy  colour  and  type  (e.g.,  Schechter
\cite{Sche76}; Lilly et al. \cite{Lill95}).  For simplicity and in order to
perform a comparative study with the C04A results, we have assumed the same
value of $M^{\star}_B=-21$ as adopted by C04A.

According  to the  synthetic templates,  we estimate  a  mean $B$-band
absolute magnitude of  $M_B=-16.5$, similar to that of  the SMC.  This
value is virtually independent on  the assumed extinction law, IMF and
metallicity.  The $M_B=-16.5$ value  corresponds to a luminosity of $L
\sim  0.016  L^{\star}$.  We  conclude  that  the  GRB~030329 host  is
clearly  a  subluminous galaxy.   This  is  consistent  with the  host
luminosity inferred by Fruchter et al. (\cite{Fruc03}).

\subsection{Estimating $A_{\rm v}$}

Among the 8 possible extinction law and IMF  combinations, we find for
50\% of the cases an extinction value within the $0.6 < A_{\rm v} \leq
0.7$  bin.    For  these  four  combinations no clear preferential
extinction law or IMF is found.  The rest of the inferred $A_{\rm v}$
values are equally distributed  in the $0.7  < A_{\rm v} \leq 0.8$ and
$A_{\rm v} \leq  0.6$ regions.  The  mean extinction value derived for
the synthetic templates is $A_{\rm v}=0.6$.

We  have  also used  the  empirical templates  by  K96  to verify  the
extinction value obtained with  the synthetic templates.  The best fit
is   then  obtained   with  the   blue   Stb1  and   Stb2  SEDs   (see
Table~\ref{table3},  $\chi^{2}/d.o.f \sim  3$), which  are  defined to
have the  lowest extinctions among the  K96 Stb sample ($0  < E(B-V) <
0.21$).  Stb templates with larger $E(B-V)$ values clearly yield worse
results.  To translate  the value of $E(B-V)$ to  $A_{\rm v}$, we have
adopted a  value of $A_{\rm  v}/E(B-V)=3.1$.  Thus, the Stb1  and Stb2
templates imply a  value of $0 < A_{\rm v} <  0.65$, in agreement with
the mean $A_{\rm  v}= 0.6$ inferred based on  the synthetic templates.
The assumed $A_{\rm v}/E(B-V)$ fraction is indicative (the actual
  $A_{\rm v}/E(B-V)$ value depends on the unknown host extinction law)
  and  only aimed to  check the  $A_{\rm v}$  value inferred  with the
  synthetic templates.

\begin{figure}[t]
\begin{center}
\resizebox{\hsize}{!}{\includegraphics[bb= 45 45 527 455]{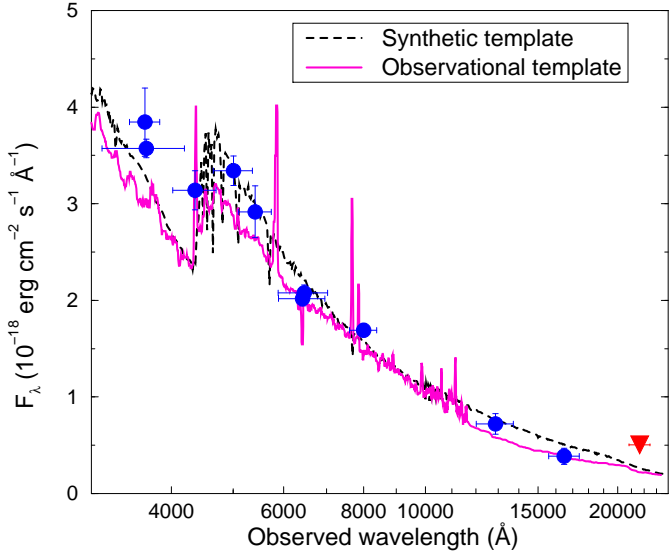}}
\caption{\label{fig2}  The  optical/NIR  SED  of the  GRB~030329  host
  galaxy.  The  circles show  the $UBVR$, $C1$,  $C2$, $C3$,  $C4$ and
  $JH$-band   detections.    The   filled  triangle   represents   the
  $K^{\prime}$  upper  limit.   The   abscissa  is  represented  on  a
  logarithmic  scale  in  order  to  enlarge  the  region  around  the
  4000~\AA\ break.  The dashed line  shows the fit obtained assuming a
  metallicity  of  $Z=0.004$, a  S55  type IMF  and  a  Prevot et  al.
  (\cite{Prev84}) extinction  law.  Note the similarity  with the Stb1
  K96 empirical template overplotted with a continuous line.  }
\end{center}
\end{figure}

Figure~\ref{fig2}  shows  the   similarities  of  the  Stb1  empirical
template and  the solution obtained  by an evolutionary  SED.  Leaving
$E(B-V)$ as  a free parameter for  the non Stb  templates, we obtained
the $E(B-V)$ values reported in column 3 of Table~\ref{table3}.  These
fits  are  not satisfactory.   Higher  values  of  $E(B-V)$ for  these
templates would make  the spectrum even redder and  hence provide even
larger  $\chi^{2}/d.o.f.$ values.   The exception  is the  Sc template
which presents a value of $\chi^{2}/d.o.f.=3.47$, just slightly larger
than for the  Stb1 and Stb2 templates.  The  $E(B-V)$ inferred for the
Sc  template is  actually  consistent  with the  one  derived for  the
empirical Stb1  and Stb2 templates, as  well as with  that derived for
the synthetic SEDs.

In principle, the  relative fluxes of the Balmer lines  can be used as
an  indicator of  the intrinsic  host galaxy  $A_{\rm  v}$ (Osterbrock
1989).  The H${\alpha}$ emission line is affected by a nearby telluric
band, so  we have  used the rest  of the  Balmer lines present  in our
spectrum  (H${\beta}$, H${\gamma}$,  H${\delta}$) to  estimate $A_{\rm
v}$. A major difficulty in using the faint Balmer lines comes from the
potential  contribution of an  underlying evolved  stellar population.
This can introduce absorption features with equivalent widths (EWs) up
to a few \AA\ (Bruzual \& Charlot \cite{Bruz03}), which are comparable
to our  measured emission fluxes  in H${\gamma}$ and  H${\delta}$.  We
have averaged the H${\gamma}$ and  H${\delta}$ absorption EWs of the 8
galaxy templates  fitted to our  photometric points and  corrected the
emission lines  fluxes to estimate $A_{\rm v}$.   The mean H${\gamma}$
and  H${\delta}$  restframe   absorption  EWs  were  $8\pm2$~\AA~  and
$7\pm3$~\AA, respectively.  The extinction values inferred are $A_{\rm
v}=0.33\pm0.19$ and $A_{\rm  v}=0.86\pm0.22$, based on H${\gamma}$ and
H${\delta}$, respectively.   The weighted  mean of the  two extinction
values is $A_{\rm v}=0.56\pm0.14$, consistent with the $A_{\rm v}=0.6$
value previously inferred based  on the synthetic and empirical galaxy
templates.   We note that  the $A_{\rm  v}$ values  determined through
H${\gamma}$,  and  especially   from  H${\delta}$,  are  significantly
sensitive to the correction for the underlying population.  Therefore,
although the  agreement is  very good, we  will assume  the extinction
determined with  the synthetic SED  templates and use  $A_{\rm v}=0.6$
throughout the paper.

\begin{table}
\begin{center}
\caption{Emission  lines  detected  in  the  GRB  030329  host  galaxy  VLT
  spectrum. The fluxes are corrected for Galactic extinction.}
\begin{tabular}{lccc}
\hline
\hline
Element       & Flux                    & Observed        & Redshift \\
              & ($10^{-17}$ erg cm$^{-2}$ s$^{-1}$)& wavelength (\AA)&\\
\hline             
$[\ion{O}{ii}]  $&$20.40\pm0.39$&$4354.08$&$0.1681$\\
$[\ion{Ne}{iii}]$&$ 4.04\pm0.28$&$4520.15$&$0.1682$\\
H${\delta}     $&$ 2.33\pm0.20$&$4792.96$&$0.1684$\\
H${\gamma}     $&$ 4.82\pm0.20$&$5071.08$&$0.1683$\\
H${\beta}      $&$11.70\pm0.36$&$5679.70$&$0.1683$\\
$[\ion{O}{iii}] $&$14.76\pm0.38$&$5793.80$&$0.1684$\\
$[\ion{O}{iii}] $&$42.21\pm0.31$&$5849.79$&$0.1684$\\
H${\alpha}     $&$32.40\pm0.47$&$7667.55$&$0.1683$\\
$[\ion{S}{ii}]  $&$ 3.27\pm0.32$&$7861.65$&$0.1684$\\
\hline

\label{table1c}
\end{tabular}
\end{center}
\end{table}

\subsection{Estimate of the metallicity}
\label{metallicity}

The  detections of [\ion{O}{ii}],   [\ion{O}{iii}]  and H${\beta}$  in
emission allow  us  to  estimate  the host   metallicity based on  the
$R_{23}$ iterative technique  (e.g., Kewley  \& Dopita \cite{Kewl02}).
Correcting the emission line fluxes for $A_{\rm v}=0.6$ and assuming a
SMC extinction law, we    obtain a metallicity of  $Z=0.004$  ($Z=0.20
Z_{\odot}$,  where $Z_{\odot}=0.02$)   and   an  ionization  parameter
$q=6.84 \times 10^7$ cm s$^{-1}$.    The inferred metallicity is  well
below  $Z_{\odot}$,   implying  that  the  GRB~030329 host   is  a low
metallicity galaxy.

The inferred metallicity  is basically insensitive to the assumed
extinction law.  Typical MW and LMC extinction laws (Pei \cite{Pei92})
yield metallicity values only $\sim  3\%$ appart from the  metallicity
derived based on the SMC.

 Hjorth et  al. (\cite{Hjor03}) reported a  rough metallicity value of
 [O/H]$=-1.0$ (equivalent  to $Z=0.10  Z_{\odot}$), not correcting for
 Galactic reddening,  intrinsic host  galaxy reddening  and underlying
 population.  If the above mentioned   corrections are not applied  to
 our measurements,  we  obtain  a metallicity  value   of [O/H]$=-0.8$
 ($Z=0.16 Z_{\odot}$).  Given that the typical metallicity uncertainty
 of   the  $R_{23}$  technique   is  $\pm0.2$~dex  (Kewley \&   Dopita
 \cite{Kewl02}),  we conclude that  our   metallicity estimate is   in
 agreement with the value inferred by Hjorth et al. (\cite{Hjor03}).

   The $R_{23}$ diagnostic  is bivaluated (lower and upper branch),
   providing two possible metallicity values.  In this study we assume
   the  metallicity  inferred  based  on the  $R_{23}$  lower  branch.
   According to  Sollerman et al.  (\cite{Soll05})  the $R_{23}$ upper
   branch   yields  a   metallicity  value   of   [O/H]$=-0.1$.   This
   metallicity degeneracy can be  broken by considering other emission
   lines such  as $[\ion{N}{ii}]$  which was not  detected in  our VLT
   spectrum.   As  noted by  Sollerman  et  al.  (\cite{Soll05}),  the
   $[\ion{N}{ii}]$ flux  upper limit derived from the  VLT spectrum is
   not constraining enough to resolve the $R_{23}$ degeneracy.

   However, the $R_{23}$ upper  branch predicts a rather bright galaxy
   ($M_B \sim -19$; Lee  et al. \cite{Lee03}),  while the lower branch
   metallicity is more consistent with the $M_B$ magnitude of the host
   ($M_B \sim -16$ is expected for $Z=0.20 Z_{\odot}$), as pointed out
   by Sollerman et al.  (\cite{Soll05}).   Therefore, the lower branch
   solution is more likely than the  upper one, but a host metallicity
   close to solar abundance can not be completely excluded.  

\subsection{Estimate of the SFR}
\label{sfroii}

Following  the  standard   H${\alpha}$  and [\ion{O}{ii}]   diagnostic
techniques (Kennicutt \cite{Kenn98},   hereafter K98) the SFR  can  be
estimated.  However,  given the contamination that H${\alpha}$ suffers
from  telluric lines, we have used  the SFR given by the [\ion{O}{ii}]
diagnostic.  This  gives a  SFR=$0.26~M_{\odot}$~yr$^{-1}$  (corrected
for  Galactic  extinction  but not for    host galaxy extinction),  in
agreement with  the  SFR derived  by  Hjorth  et  al.  (\cite{Hjor03};
SFR=$0.20\pm0.07~M_{\odot}$~yr$^{-1}$   once  it    is corrected   for
Galactic extinction).  For  a Pei  (\cite{Pei92}) SMC-like  extinction
law, and  adopting $A_{\rm  v}=0.6$, we  obtain an unextincted  SFR of
$0.6~M_{\odot}$~yr$^{-1}$.

A verification of the SFR inferred  from the [\ion{O}{ii}] line can be
carried   out  estimating three independent  SFRs  through H${\beta}$,
H${\delta}$, and H${\gamma}$ assuming  a typical case B  recombination
and taking account the  extinction  previously derived.  The  inferred
three SFRs show  a mean  weighted value of  $0.63~M_{\odot}$~yr$^{-1}$
and  a  dispersion   of  $0.15~M_{\odot}$~yr$^{-1}$.  This   is  fully
consistent with the SFR derived based on [\ion{O}{ii}].

The photometric  SEDs bring an additional  diagnostic to  estimate the
SFR.   The method for deriving the  SFR from the  UV continuum flux is
one of  several diagnostic methods used  in  the literature to measure
SFRs in galaxies (see K98 for a  comprehensive review).  Obviously, if
there is  dust-enshrouded star  formation,  then this  UV-based method
will only provide  a lower limit to  the actual SFR.  At $z=0.168$ the
2800~\AA\ region is redshifted to $\sim$3270~\AA, so it is included in
the sensitivity  region of the $U$-band and  the BUSCA  $C1$ band.  We
have  derived the average    value  of the   fluxes  in  both  filters
($F_{2800{\rm  \AA}}\sim  1.5\times10^{-29}$   erg s$^{-1}$  Hz$^{-1}$
cm$^{-2}$) and determined the  UV luminosity at restframe 2800~\AA\ to
be $L_{\rm     UV}\sim1.2 \times  10^{27}$   erg  s$^{-1}$  Hz$^{-1}$.
Following  K98  this corresponds  to a SFR=0.17~$M_{\odot}$~yr$^{-1}$.
Given  that the  rest frame   luminosity of the  host   is $L =  0.016
L^{\star}$,  we obtain  a specific   star   formation (SSFR) rate   of
SSFR=10.6~$M_{\odot}$ yr$^{-1} (L/L^{\star})^{-1}$.  If we correct the
previous UV SFR  and SSFR estimates for   the reddening effect  in the
host galaxy  ($A_{\rm v}\sim 0.6$, assuming a  SMC extinction law), we
obtain unextincted  SFR $\sim  0.54~M_{\odot}$~yr$^{-1}$ and SSFR$\sim
34~M_{\odot}$ yr$^{-1} (L/L^{\star})^{-1}$.  Thus, the unextincted SFR
derived from the UV continuum is in agreement with the unextincted SFR
inferred based  on the [\ion{O}{ii}] and  the Balmer lines  (SFR$ \sim
0.6 M_{\odot}$~yr$^{-1}$).   Given that C04A   used the  UV diagnostic
technique for their   SSFR  estimates,  for comparative   purposes  we
adopted the SSFR  derived from the  UV continuum  ($\sim 34~M_{\odot}$
yr$^{-1} (L/L^{\star})^{-1}$).

\subsection{A potential galaxy association at $z\sim 0.17$?}

A  GRB\,030329  field galaxy   located  at $\alpha_{J2000}=  10^h 45^m
01.46^s$, $\delta_{J2000}= +21^{\circ} 29^{\prime} 30\farcs1$ (with an
internal error of $0\farcs7$  per coordinate) was accidentally  placed
on the FORS2 slit.  The galaxy showed prominent emission lines similar
to the host    galaxy,  and turned  out   to  be  at   a redshift   of
$z=0.1710\pm0.0003$.  The similarity to  the GRB host galaxy redshift,
naturally opens the question  on a potential clustering or association
at $z\sim  0.17$. The nearby galaxy is  located 192.4  arcsec from the
host,   implying  a  distance  of  $\sim600$  kpc at  the  host galaxy
redshift. The radii of galaxy  clusters is of the order  of a few Mpc,
so the measured angular distance to the host galaxy is compatible with
a  physical   association at $z\sim 0.17$.     On the other  hand, the
velocity difference  is $\sim 685  \pm 104$ km s$^{-1}$,  in agreement
with the velocity dispersions   measured  in $z\lesssim 0.15$   galaxy
clusters (Desai et al.  \cite{Desa04}).

A spectroscopic study of a significant  galaxy sample in the GRB field
would   be necessary to investigate   whether  the host is  physically
linked to other angularly  close galaxies.  If  such an association is
confirmed, the case of the GRB\, 030329 host would  be opposite to the
preferential  tendency of GRB  hosts to reside  in  low galaxy density
environments (Bornancini et al. \cite{Born04}).

\subsection{Galaxy type}
\label{type}

\begin{table}
\begin{center}
\caption{\label{table3}  Main  results  obtained  based  on  the  empirical
  templates  by Kinney et  al.  (\cite{Kinn96}).   The templates  have been
  ordered according to their colour, bluer  on the top (Stb1) and redder at
  the bottom (B).  There is a clear correlation between the goodness of the
  fit  and the  colour of  the  template: the  bluer the  colour the  lower
  $\chi^2/d.o.f.$.}
\begin{tabular}{lccc}
\hline
\hline
Galaxy   &$\chi^2/d.o.f.$&$ E(B-V)$ \\
Template &($d.o.f. = 9$) &         \\
\hline
Stb1    &$3.06$  & $0.05\pm0.05^{\dagger}$ \\ 
Stb2    &$3.01$  & $0.15\pm0.05^{\dagger}$ \\ 
Stb3    &$14.86$ & $0.30\pm0.05^{\dagger}$ \\ 
Stb4    &$24.48$ & $0.45\pm0.05^{\dagger}$ \\ 
Stb5    &$36.77$ & $0.55\pm0.05^{\dagger}$ \\ 
Stb6    &$47.57$ & $0.65\pm0.05^{\dagger}$ \\ 
Sc      &$3.47$  & $0.18$ \\ 
Sb      &$133.38$& $0$  \\ 
Sa      &$153.72$& $0$  \\ 
S0      &$176.87$& $0$  \\ 
E       &$182.18$& $0$  \\ 
B       &$180.96$& $0$  \\ 
\hline
\multicolumn{3}{l}{$\dagger$ Extinction fixed by the template definition.}\\
\hline
\end{tabular}
\end{center}
\end{table}

When using the synthetic templates for  the fit, we found that the favoured
galaxy type  is not very dependent  on the assumed IMF  and extinction law.
Among the 8 IMF/extinction law combinations, a starburst galaxy is obtained
in 50\% of the  cases.  The rest of cases are distributed  in Sa (25\%), S0
(12.5\%) and Sb (12.5\%).  Thus the synthetic SED fits favour a Stb type.

This  result is  supported by  the  fits obtained  with the  empirical
templates  by  K96  (Table~\ref{table3}).   In general  the  starburst
templates give lower values  for $\chi^2/d.o.f.$ when fitting the SED.
The only exception is for  the Sc type, $\chi^2/d.o.f.$=$3.47$ , whose
value is quite  close to the best ones obtained for  the Stb1 and Stb2
templates.

\subsection{Dominant stellar population age}
\label{age}

The inferred mean stellar population  age for the 8 IMF/extinction law
combinations is  150$\pm$80 Myr.  In principle  progenitors older than
$\sim$50  Myr are  not  obvious to  accommodate  within the  collapsar
scenario, since the  age of an $\sim8 M_{\odot}$  star (the minimum SN
progenitor mass)  when it explodes  as a core-collapse SN  is $\sim$50
Myr (see for instance Portinari et al. \cite{Port98}).

However,  an  ideal starburst  where  all  the  star formation  occurs
simultaneously, and hence all the  stars have exactly the same age, is
not realistic.  Thus, in a  real starburst, like the  GRB~030329 host,
although  the majority of  the star  formation occurred  $\sim150$ Myr
ago, the  star formation can not  stop sharply, but has  a stellar age
dispersion  which  spreads  down  to  $150$ Myr.   Hence,  an  age  of
$\sim150$ Myr  should be considered  an upper limit of  the progenitor
age.

An  independent  check  of  the  dominant stellar  population  age  can  be
performed comparing the restframe H${\beta}$ equivalent width derived from
the VLT  spectrum (EW$\sim$50~\AA, corrected for  the underlying population)
to   the   one  predicted   by   Starburst99   (SB99,   Leitherer  et   al.
\cite{Leit99}).   According to  SB99, for  a metallicity  of  $Z=0.004$ our
H${\beta}$ EW corresponds to ages  ranging from $\sim10$ Myr (for the case
of an ideal instantaneous SFR, $\tau = 0$) to $\sim300$ Myr (for a constant
SFR, $\tau =  \infty$).  Our age of $150\pm80$ Myr  is consistent with this
age  range, supporting  that the  star formation  might be  brief,  but not
perfectly instantaneous ($\tau = 0$).

\subsection{Host galaxy stellar mass}
\label{mass}

 Given that the Bruzual \& Charlot (\cite{Bruz03}) templates are normalized
 to one Solar mass, the normalization factor used to scale the templates to
 our photometric fluxes allowed us to estimate the host galaxy stellar mass
 ($M_{\rm host}$).  We caution  however that the optical photometric points
 are likely dominated by young massive stars, so this estimator tends to be
 insensitive  to the  redder  low-luminosity (but  very abundant)  low-mass
 underlying  stellar population.   This  effect is  especially relevant  in
 galaxies, like the GRB\, 030329 host, where the optical light is dominated
 by a bright starburst episode(s).  So, strictly speaking, the derived mass
 should be  considered a lower limit  to the total stellar  mass present in
 our host.

 Assuming the  S55 and CH03 IMFs and  not accounting for the  effect of the
 internal extinction, we  derived mass values of $M_{\rm  host} = 9.1\pm1.3
 \times  10^{7} M_{\odot}$  and  $M_{\rm host}  =  6.0\pm0.9 \times  10^{7}
 M_{\odot}$,  respectively.   Given that  an  extinction  value of  $A_{\rm
 v}=0.6$ implies  a scale up of the  intrinsic optical flux by  a factor of
 $\sim2$, we adopt $10^{8} M_{\odot}$  as a conservative lower limit to the
 total stellar mass.

  An independent rough estimation of this value can be obtained through the
  statistical mass-to-luminosity  ratios ($M/L$) reported  for the galaxies
  of the  Sloan Digitized  Sky Survey (SDSS;  York et  al.  \cite{York00}).
  Adopting   $K=K^{\prime}$,   for    the   $K$-band   absolute   magnitude
  ($M_{K}=-19.1$, basically  not affected by  the host galaxy  reddening if
  $A_{\rm v} \sim  0.6$ is assumed) and the $B-R$  restframe colour of our
  host,  $M/L  \sim  0.7$  is  expected   (see  Table  7  of  Bell  et  al.
  \cite{Bell03}).  This  would imply a  stellar mass of $M_{\rm  host} \sim
  7\times 10^{8} M_{\odot}$.  The $M/L$-colour relations derived by Bell et
  al.   (\cite{Bell03}) are based  on a  large sample  of galaxies  with no
  distinction of the galaxy type, so  they represent an average over a wide
  morphological range. However, for  a given colour, low-mass galaxies tend
  to  show lower  $M/L$  values (Binney  \&  Merrifield \cite{Binn98}),  so
  $7\times 10^{8} M_{\odot}$ is a  plausible upper limit to the host galaxy
  stellar mass.  Therefore, we conclude  that the total stellar mass of our
  galaxy should be $10^{8} M_{\odot} \lesssim M_{\rm host} \lesssim 7\times
  10^{8} M_{\odot}$.

\section{Discussion}
\label{Discussion}

 The results above were  obtained assuming the $R_{23}$ low metallicity
branch (see  Sollerman et al.   \cite{Soll05}). To verify that  our results
are not  critically dependent  on this assumption  we have also  run models
with Solar abundances.

 No remarkable differences were found in the output parameters.  This is in
 agreement  with  the  extensive  tests  performed  by  Bolzonella  et  al.
 (\cite{Bolz00}) who reported that the metallicity is a secondary parameter
 for the SED fitting technique

 The most relevant  impact of the metallicity is the  effect on the stellar
 age.  Solar  metallicity templates show  slightly redder colours  than the
 $Z=0.004$ templates,  therefore shorter time evolution is  required to fit
 the  photometric points, yielding  younger stellar  ages.  Thus,  the mean
 stellar  population age  inferred  using Solar  metallicity templates  was
 $\sim 43 \pm 7$  Myr, a factor three lower than the  one inferred with the
 $Z=0.004$ templates.
 
 The metallicity  showed a second  order impact on $A_{\rm  v}$. Given
 that the  Solar metallicity templates are redder,  they provide lower
 $A_{\rm v}$  values.  Thus, assuming  Solar metallicity we  obtain an
 extinction  of  $A_{\rm  v}=0.4$  (instead  of  $A_{\rm  v}=0.6$  for
 $Z=0.004$), yielding  an UV unextincted SFR=$0.37~M_{\odot}$yr$^{-1}$
 and  SSFR=$23.1~M_{\odot}$ yr$^{-1}$ ($L/L^{\star})^{-1}$.   The rest
 of the  parameters (photometric redshift, galaxy  type, stellar mass)
 are  basically  insensitive  to  the  metallicity  of  the  synthetic
 templates.

\subsection{The GRB~030329 host SSFR relative to galaxies at a similar redshift}
\label{sdss}

We  performed a  comparative  analysis  of the  GRB~030329  host SSFR  with
respect to  the SSFRs obtained for a  galaxy sample based on  the SDSS.  In
order to  minimize any impact  that the SSFR--redshift evolution  (Cowie et
al.   \cite{Cowi96};  Juneau  et  al.   \cite{June05}) might  have  on  our
analysis,  only  galaxies  with  0.160$  <  z  <  $0.175  were  considered.
Following the process  explained in C04A, the unextincted  SSFR of the SDSS
galaxies were calculated using HyperZ.   We then selected only the galaxies
that showed acceptable SED fits ($\chi^2/d.o.f \leq$ 1.5).  With the aim of
estimating the  probability errors,  the process was  repeated for  the two
IMFs given by S55 and CH03.

Assuming a S55  IMF, among the 12333 SSFRs calculated  for the SDSS sample,
only    753   showed    SSFRs   larger    than    $34~M_{\odot}$   yr$^{-1}
(L/L^{\star})^{-1}$, implying  only 6.1\% of  the sample.  The  results did
not  change qualitatively when  the CH03  IMF was  used (6.9\%).   Hence we
estimate that the GRB\, 030329 host SSFR is higher than $\sim$93.5\% of the
galaxies at an equivalent redshift.

This  result  is  also  supported  by an  independent  (but  more  reduced)
comparison sample based on 1067+1611 HDF galaxies (adding the catalogues by
Fern{\'   a}ndez-Soto   et  al.    \cite{Fern99}   and   Vanzella  et   al.
\cite{Vanz01}).  For  a S55  IMF, only one  of the  89 HDF galaxies  in the
0.05$ < z  <$ 0.25 redshift slot showed  a SSFR larger than the  one of the
GRB\, 030329  host (98.9\%  percentile).  If  a redshift 0$ < z <$ 0.25
window is  assumed (containing  143 galaxies), then  the GRB\,  030329 host
SSFR  is  over  97.9\%  of  the sample.   Again  the  conclusions  remain
basically invariant  for a CH03 IMF  (in a 0$  < z < $0.25  redshift window
only  5  among  136  galaxies  displayed  SSFR  $>  34~M_{\odot}$  yr$^{-1}
(L/L^{\star})^{-1}$; 96.3\% percentile).

\subsection{The GRB~030329 host in the framework of previous compilations}

The SFR  (corrected for intrinsic extinction) of  the GRB\,030329 host
is the lowest among the C04A sample. However, the SSFR is the highest,
given that  the $M_B$ of  our host is  1.6 magnitudes dimmer  than the
faintest  C04A sample  galaxy.  The  GRB\,030329 host  falls  into the
selection  criteria  defined by  C04A  ($R  <$  25.3, known  redshift,
detection in  5 or more  filters, not very complex  morphology).  This
allowed us to  incorporate our host to the C04A  10 host galaxy sample
and repeat part of the C04A statistical analysis.

In  order  to  construct  a  reference  sample  with  similar  spatial
distribution to the (10+1) hosts,  we calculated the SSFR only for the
HDF galaxies with redshifts consistent with our 11 host galaxies (0.17
$\leq z \leq$  2.1).  Selecting only HDF galaxies  with acceptable SED
fits ($\chi^2/d.o.f \leq $1.5) reduced  the HDF catalogues to 1917 for
a S55 IMF, and 1880 for a CH03 IMF.

The SSFR  distribution of our 11  hosts and the  HDF reference samples
were  statistically compared  using the  two-sample Kolmogorov-Smirnov
test  (Peacock  \cite{Peac83};  see   C04A  for  more  details).   The
probability  that   the  two  samples  share  the   same  parent  SSFR
distribution   ranges   from   $p=1.2\times10^{-4}$   (S55)   to   $p=
5.5\times10^{-4}$   (CH03)  depending  on   the  assumed   IMF.   This
calculation reduces the  C04A $p$ values by an  order of magnitude and
strengthens the  previous C04A results,  who concluded that  GRB hosts
show on average higher SSFRs than HDF galaxies at similar redshifts.

 According  to our synthetic SED fitting  a $K^{\prime}$-band magnitude
 of $K^{\prime}\sim  $20.5 is predicted, implying $R-K  \sim$ 2.1 (adopting
 $K =  K^{\prime}$ and  correcting for Galactic  reddening). This  colour is
 consistent with the colours of the  host galaxy sample by Le Floc'h et al.
 (\cite{LeFl03}).  The  $K$-band absolute magnitude estimated  for the host
 ($M_K=-19.1$)  is in agreement with  the general results by  Le Floc'h et
 al. (\cite{LeFl03}) who noted that GRB hosts are significantly subluminous
 in the $K$-band (compare to their Fig.~5).

 The  estimated $R-K$ colour and the  low redshift of the GRB\, 030329
 host galaxy  are  consistent with the  conclusions  by  Berger et al.
 (\cite{Berg03}) and Tanvir et  al.   (\cite{Tanv04}), who found  that
 the detected GRB hosts  tend to be  bluer and at lower redshifts than
 the submm-selected galaxies.    In particular, it is illustrative  of
 the position  of the GRB\,  030329 host in the Fig.~6 colour-redshift
 diagram by Berger et al.  (\cite{Berg03}).

 On the  other hand, the estimated  host galaxy stellar mass  is lower than
 the ones calculated  for the host sample of  Chary et al. (\cite{Char02}),
 may  be with  the exception  of the  GRB\,970508 host  galaxy  (Sokolov et
 al. \cite{Soko01}).

\section{Conclusions}
\label{Conclusions}

VLT spectroscopic observations  show that the GRB\,030329 host  is a likely
low metallicity galaxy, with $Z=0.004$ ($Z\sim Z_{\odot}/5$).  From the SED
fitting we infer that the host  of GRB\,030329 is most probably a starburst
galaxy.   This  agrees  with the  C04A  results,  who  found (based  on  an
independent host sample) that GRB  hosts correspond to Stb galaxies in 90\%
of the cases.   We derived a dominant stellar  population age of $150\pm80$
years,  in agreement with  the stellar  age distribution  of the  C04A host
galaxy sample (Mean age = 95  $\pm$ 74 Myr).  The low absolute magnitude of
the  host,  $M_B \sim  -16.5$  ($L  \sim  0.016 L^{\star}$),  confirms  its
subluminous nature.

The SED fitting  yields a consistent extinction value of  $A_{\rm v} = 0.6$
using  either  synthetic  or   empirical  templates.   This  extinction  is
consistent with  the mean extinction derived  by C04A (median  $A_{\rm v} =
0.26^{+0.51}_{-0.26}$).   The extinction derived  from the  Balmer emission
lines is also consistent with the one derived from the SED.

We  determine   a  consistent  unextincted   SFR$\sim  0.6  M_{\odot}$
yr$^{-1}$  based on  three independent  diagnostic techniques;  the UV
continuum,  the [\ion{O}{ii}],  and the  Balmer emission  lines.  This
consistency  supports  the inferred  $A_{\rm  v}  =  0.6$ value.   The
associated   unextincted   SSFR    ($\sim   34   M_{\odot}$   yr$^{-1}
(L/L^{\star})^{-1}$) is higher than  $\sim93.5\%$ of the SDSS galaxies
at  $z\sim0.17$.    Moreover,  a  statistical   analysis  carried  out
enhancing the  C04A sample with  the GRB\, 0303029 host  confirms that
GRB hosts shows  higher SSFRs than HDF galaxies  at similar redshifts.
All these findings are consistent with the host galaxy being an active
star forming galaxy.

We note the presence of a  close galaxy with a similar redshift to the
GRB  host,  which  may  be  indicative of  a  possible  clustering  at
$z\sim$0.17.  Multi-object  spectroscopy of the  GRB\,030329 field may
clarify this open issue.

\begin{figure}[t]
\begin{center}
\resizebox{\hsize}{!}{\includegraphics[bb= 40 10 578 405]{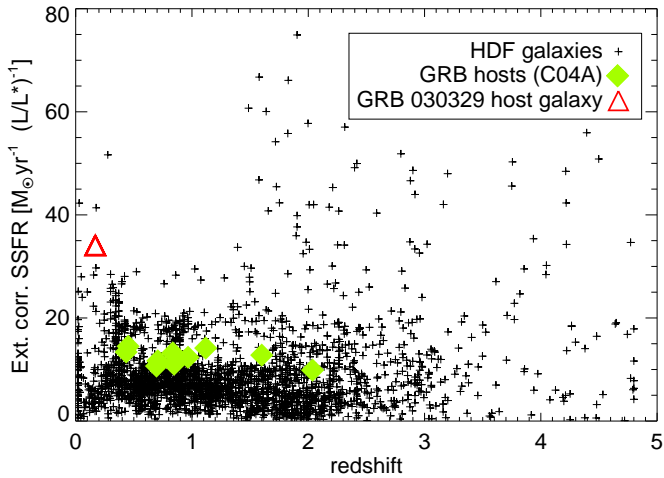}}
  \caption{\label{ssfr}  The  unextincted  SSFR  of  the  C04A  sample
  (filled diamonds)  plus the GRB~030329  host galaxy (triangle)  as a
  function of  redshift.  The crosses show  the SSFR of  the HDF
  when a  S55 IMF is assumed.   Note the high SSFR  of the GRB\,030329
  host in comparison to the HDF galaxies at a similar redshift.}
\end{center}
\end{figure}

\section*{Acknowledgments}

JG acknowledges the  support of a Ram\'on y  Cajal Fellowship from the
Spanish Ministry  of Education and  Science.  The research of  DPR has
been supported  by the Education  Council of Junta  de Andaluc\'{\i}a,
Spain. This research is partially supported by the Spanish Ministry of
Science  and  Education  through programmes  ESP2002-04124-C03-01  and
AYA2004-01515 (including FEDER  funds).  The observations presented in
this   paper  were   partially  obtained   under  the   ESO  Programme
271.D-5006(A).  PJ,  GJ and GB  gratefully acknowledge support  from a
special grant from  the Icelandic Research Council.  VVS  and TAF were
supported  by the  Russian  Foundation for  Basic  Research, grant  No
01-02-171061.  We thank S. Savaglio for useful comments and C.  Kehrig
for  helpful   assistance  analysing   the  VLT  spectrum.    We  also
acknowledge S. Toft, J.-M.  Miralles, M.  Bolzonella for their support
with HyperZ and our anonymous referee for constructive comments.

\end{document}